\documentclass[fleqn,twoside]{article}
\usepackage{espcrc2}

\usepackage{graphicx}
\usepackage[figuresright]{rotating}

\newcommand{\AmS}{{\protect\the\textfont2
  A\kern-.1667em\lower.5ex\hbox{M}\kern-.125emS}}

\usepackage{amsmath}
\usepackage{amssymb}

\title{Quark mass dependence of nucleon mass and axial-vector coupling constant}

\author{M. Procura\address[TU]{Physik-Department, Theoretische Physik\\ Technische Universit\"at M\"unchen, D-85747 Garching, Germany}, %
        T.R. Hemmert\addressmark[TU], 
        B.U. Musch\addressmark[TU]
	and
        W. Weise\addressmark[TU]}       
\begin{document}

\begin{abstract}
We present an updated analysis of the quark mass dependence of the nucleon mass $M_N$ and nucleon axial-vector coupling $g_A$, comparing different formulations of $SU(2)$ Baryon Chiral Effective Field Theory, with and without explicit $\Delta\,(1232)$ degrees of freedom. We discuss the outcome of the corresponding interpolations between lattice QCD data and the physical values for these two nucleon observables. It turns out that in order to obtain successful interpolating functions at one-loop order, the inclusion of explicit $\Delta\,(1232)$ degrees of freedom is not decisive for the nucleon mass but crucial for $g_A$. A chiral extrapolation of recent lattice results by the LHP collaborations is also shown.
\vspace{1pc}
\end{abstract}

\maketitle

\section{THE NUCLEON MASS}

In Ref.\cite{PHW} we worked out the quark mass dependence of $M_N$ at the leading and next-to-leading one-loop level in manifestly covariant $SU(2)$ Baryon Chiral Perturbation Theory (BChPT). We neglect isospin breaking effects and translate the functional dependence on the (light) quark mass into a pion mass dependence according to the Gell-Mann-Oakes-Renner relation.
We evaluate all the relevant graphs using the so-called infrared regularization method \cite{BL} which represents a variant of dimensional regularization able to treat one-loop integrals involving baryon propagators in a way consistent with chiral power counting, in a manifestly Lorentz invariant framework.

In the numerical analysis we have estimated parameters not fixed by chiral symmetry by fitting the relevant formulae to a combined set of fully dynamical two-flavor lattice QCD data for $M_N$ versus $m_\pi$ obtained by the CP-PACS\cite{CPPACS}, JLQCD\cite{JLQCD} and QCDSF-UKQCD\cite{QCDSF,UKQCD} Collaborations. In order to minimize artifacts from discretization and finite volume effects, we have selected from the set of available simulations those with lattice spacing $a<0.15\,{\rm{fm}}$ and $m_\pi\,L>5$, where $L$ is the spatial size of the lattice. We have not considered partial quenching effects, selecting lattice calculations with valence quark masses equal to sea quark masses. We assume that there is an overlap between the region of quark masses where $SU(2)$ BChPT is applicable and the range of quark masses presently accessible to full-QCD lattice calculations, and explore the consequences of this assumption. We have restricted ourselves to $m_\pi<600\,{\rm{MeV}}$.

\begin{figure}[t]
  \begin{center}
    \includegraphics*[width=0.5\textwidth, trim=0 0 0 -10]{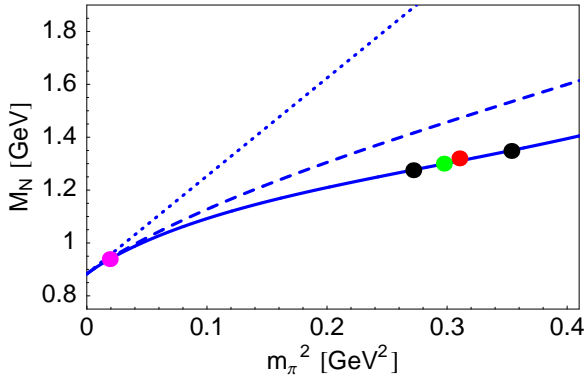}
    \caption{Solid curve: best fit using the next-to-leading one-loop ${\cal O}(p^4)$ expression for $M_N(m_\pi)$ in covariant $SU(2)$ BChPT, without explicit $\Delta\,(1232)$ degrees of freedom. The physical point is included as a constraint. Dashed curve: leading-one-loop curve obtained from the ${\cal O}(p^4)$ fit. Dotted curve: ${\cal O}(p^2)$ result.}
    \label{figconv}
  \end{center}
\end{figure}

In Fig.\ref{figconv} we plot the best-fit curve coming from the next-to-leading one-loop expression. We include the physical point as a constraint. The output parameters show high degree of consistency with information from low-energy $\pi N$ and $N N$ scattering \cite{PHW,PMHWW}. Statistically compatible results have been obtained by fitting with the ${\cal O}(p^4)$ Heavy Baryon expression, which coincides with the ${\cal O}(p^4)$ expression in the covariant framework truncated at $m_\pi^4$ in the chiral expansion. This confirms the outcome of Ref.\cite{cutoff}. Furthermore, we have taken into account finite (spatial) volume effects, and fitted to an enlarged set of lattice data ($L>1\,{\rm fm}$) \cite{PMHWW} using the ${\cal O}(p^4)$ expression worked out by the QCDSF-UKQCD collaboration \cite{QCDSF}. Our results in the infinite volume are very nicely confirmed by those fits \cite{PMHWW}.
Fig.\ref{figconv} also shows that the difference between the leading and the next-to-leading one-loop result can be kept reasonably small over the whole range of $m_\pi$ that we analyze. 

The ${\cal O}(p^4)$ pion mass dependence of the pion-nucleon sigma term can be obtained by applying the Feynman-Hellmann theorem to the expression of $M_N(m_\pi)$ at the next-to-leading one-loop order. According to the outcome of the corresponding fit for $M_N$, we obtain $\sigma_{\pi N}=49 \pm 3\,{\rm{MeV}}$ at the physical pion mass \cite{PHW}, in perfect agreement with the outcome of the analysis by Gasser, Leutwyler and Sainio, $\sigma_{\pi N}=45 \pm 8\,{\rm{MeV}}$ \cite{GLS}. 

\subsection{Statistical analysis}
\begin{figure}[t]
  \begin{center}
    \includegraphics*[width=0.5\textwidth,trim=16 0 0 0]{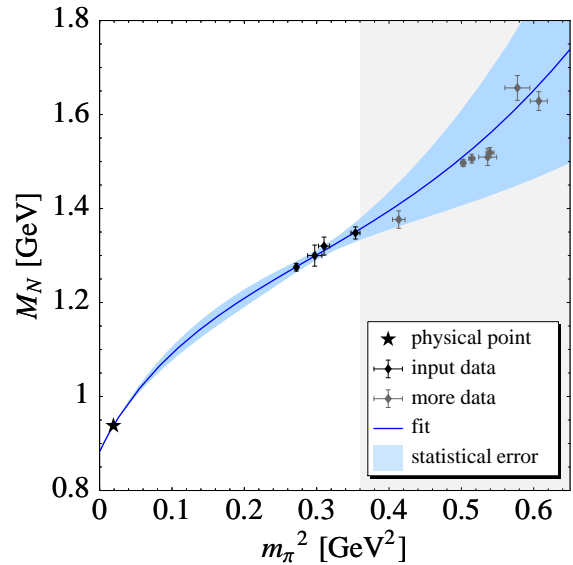}
    \caption{$M_N(m_\pi)$: global statistical error band associated with the 68\% joint confidence region for the fit parameters at order $p^4$. The physical point together with the data up to $600\,{\rm MeV}$ in $m_\pi$ have been included as input. The best-fit curve is also plotted.}
    \label{bands}
  \end{center}
\end{figure}

In Fig.\ref{bands} we show the statistical error band corresponding to the 68\% joint confidence region for the fit parameters in $M_N(m_\pi)$ at order $p^4$ \cite{PMHWW}. The errors on the lattice pion masses have been taken into account as well. The inclusion of the physical nucleon mass is crucial to shrink the bands below about $300\,{\rm MeV}$ in pion mass. In order to extract information from present lattice data about the region of small quark masses within our systematic approach, we should either incorporate phenomenological input or perform simultaneous fits to several observables characterized by a common subset of low-energy parameters.

\subsection{Including explicit $\Delta\,(1232)$ degrees of freedom}

\begin{figure}[t]
  \begin{center}
    \includegraphics*[width=0.5\textwidth, trim=16 0 0 0]{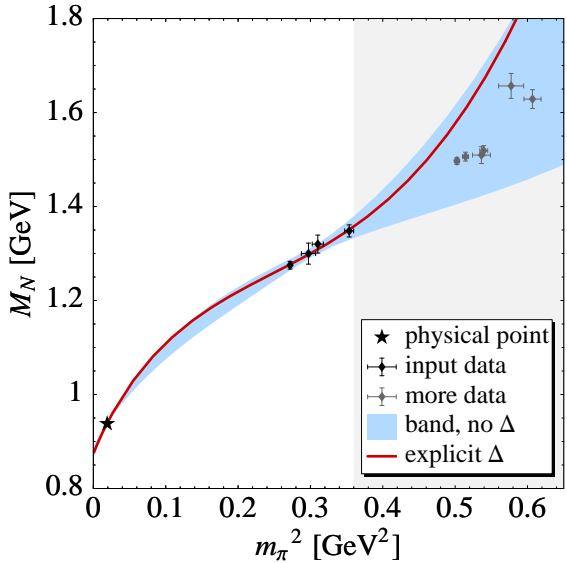}
    \caption{$M_N(m_\pi)$ with explicit $\Delta(1232)$ degrees of freedom: best-fit curve based on the leading-one-loop expression for $M_N(m_\pi)$ in manifestly covariant Small Scale Expansion. The 68\% confidence level error band for the fit at order $p^4$ in covariant $SU(2)$ BCHPT with $\pi N$ degrees of freedom is shown as well.}
    \label{deltamass}
  \end{center}
\end{figure}

Following the strategy outlined above we have analyzed $M_N(m_\pi)$ in the context of an effective field theory which apart from nucleons and pseudo-Goldstone bosons includes the $\Delta(1232)$ as an explicit degree of freedom \cite{PMHWW}. The theoretical framework of our analysis is covariant $SU(2)$ BChPT with infrared regularization in the presence of spin-3/2 fields \cite{BHM,BHMdeltadue}. The delta-nucleon mass splitting in the $SU(2)$ chiral limit is treated as a small parameter and included in the power counting according to the so-called Small Scale Expansion (SSE)\cite{SSE}.
Fig.\ref{deltamass} shows the pion mass dependence of $M_N$ at leading-one-loop order in manifestly covariant SSE \cite{PMHWW}, see Refs.\cite{BHM,BHMdeltadue}. The curve is again obtained by imposing the physical constraint and fitting to the available two-flavor fully dynamical lattice QCD data compatible with the cuts explained above, up to $m_\pi= 600\,{\rm MeV}$. As in the previous case, the estimated parameters come out of natural size and compatible with available information from phenomenology \cite{PMHWW}. In Fig.\ref{deltamass} the best-fit curve is plotted together with the 68\% error band for the ${\cal O}(p^4)$ fit in the scheme with pion and nucleon as explicit degrees of freedom. Treating the $\Delta(1232)$ as a dynamical variable is not essential for a satisfactory description of the pion mass dependence of the nucleon mass. An equally successful interpolation can be obtained by ``freezing'' the delta effects into low-energy constants.

\section{THE AXIAL\,-\,VECTOR COUPLING CONSTANT $g_A$}

We have compared also for $g_A$ the chiral effective field theory schemes with and without explicit $\Delta\,(1232)$ degrees of freedom, both in their non-relativistic \cite{HPW} and manifestly Lorentz invariant form with infrared regularization \cite{PHWga}, see also \cite{barpro}. 
We have evaluated the contributions up to leading and next-to leading one-loop order in $SU(2)$ BChPT and used as input for the numerical analysis a set of (quenched) lattice data points provided by RIKEN-BNL-Columbia-KEK Collaboration \cite{RBCK}. Full-QCD simulations for $m_\pi \le 600\,{\rm{MeV}}$ are reported by the same collaboration to be fully consistent with the quenched results \cite{RBCK2}.
In the scheme {\it without} explicit $\Delta\,(1232)$ degrees of freedom, both at leading and next-to-leading one-loop level, it is impossible to get a successful intepolating curve between lattice data and the physical point with parameters in agreement with low-energy $\pi N$ scattering analyses, cf. also the talk by Ulf-G. Mei{\ss}ner in this workshop and Ref. \cite{ulflatt}. However, intermediate spin-3/2 resonance contributions are known to play a very important role in this context. Recall, for example, the Adler-Weisberger sum rule \cite{AW} which relates the deviation of $g_A$ from 1 to pion-nucleon dynamics and spontaneous chiral symmetry breaking:
\begin{equation}
g_A^2=1+\frac{2 f_\pi^2}{\pi}\int_{0}^{\infty}\frac{d q}{\omega}\,[{{\sigma_{\pi^+ p}(\omega)}}-{{\sigma_{\pi^- p}(\omega)}}]+\,\dots \nonumber
\end{equation}
where the integral of the difference of the $\pi^+ p$ and $\pi^- p$ total cross-sections is taken over the pion momentum in the nucleon rest frame and $\omega$ denotes the pion energy in that frame. Terms of order $m_\pi^2/M_N^2$ are neglected. The $P$-wave contributions from the nucleon and delta pole terms \cite{EW} give total $\pi^{\pm} p$ cross-sections which, according to the Adler-Weisberger sum rule, yield $g_A \sim 1.24$. If we take into account only the $P$-wave Born terms for static nucleons, we get $g_A \sim 0.94$ \cite{PHWga}.

\begin{figure}[t]
  \begin{center}
    \includegraphics*[width=0.5\textwidth]{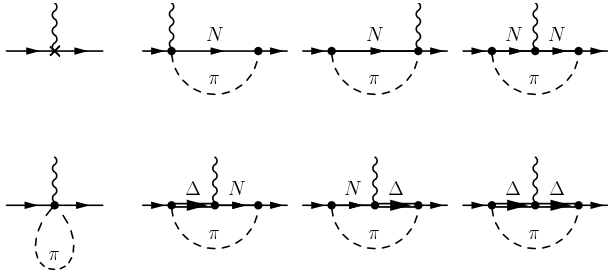}
    \caption{Diagrams contributing to the nucleon axial-vector coupling constant $g_A$ at leading-one-loop order with explicit $\Delta\,(1232)$ degrees of freedom. The wiggly line denotes an external isovector axial-vector field.}
    \label{diag}
  \end{center}
\end{figure}

\begin{figure}[h]
  \begin{center}
    \includegraphics*[width=0.5\textwidth]{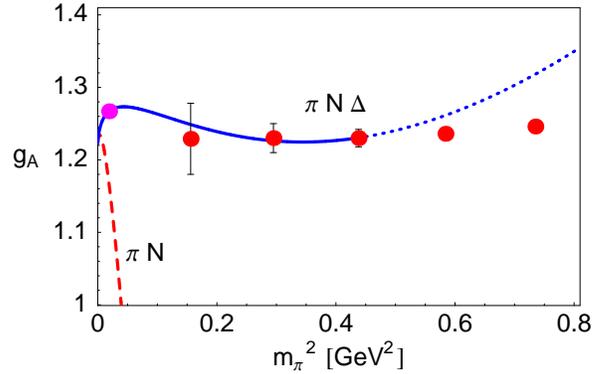}
    \caption{Effects of the inclusion of the explicit $\Delta(1232)$ degrees of freedom. The dark solid dots refer to simulations by the RBCK Collaboration \cite{RBCK}.}
    \label{gauno}
  \end{center}
\end{figure}

We have analyzed $g_A$ at the leading one-loop order in the framework with explicit $\Delta\,(1232)$ degrees of freedom (non-relativistic SSE) evaluating the graphs in Fig.\ref{diag}.
The best-fit curve including the physical constraint is shown in Fig.\ref{gauno}. The axial-$\Delta$-$\Delta$ coupling $g_1$ is treated as a free parameter, while the value of axial-$N$-$\Delta$ coupling is an input for the fit, cf. Fig.\ref{diag}. For a choice of the latter according to the relativistic expression of the $\Delta \to \pi N$ strong decay width, the fit prefers values of the former close to the $SU(4)$ quark model prediction $g_1=9/5\, g_A$. We refer to \cite{gafin} for an analysis of finite volume effects for $g_A$ in the same theoretical framework we use. See also the contribution by Meinulf G\"ockeler to this workshop. 

\begin{figure}[t]
  \begin{center}
    \includegraphics*[width=0.5\textwidth]{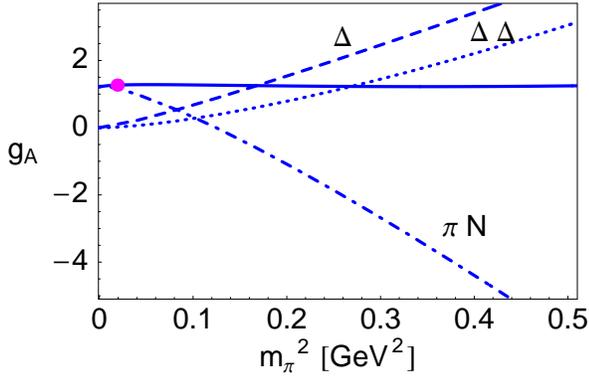}
    \caption{Solid-curve: best-fit curve at leading-one-loop in SSE, Fig.\ref{gauno}. The dot-dashed, dashed and dotted curves refer to the diagrams without, with one and with two $\Delta$ propagators in Fig.\ref{diag}, respectively.}
    \label{gadue}
  \end{center}
\end{figure}

In Fig.\ref{gauno} we plot also the leading-one-loop (dashed) curve in Heavy Baryon Chiral Perturbation Theory, as extracted from the fit in SSE, cf. \cite{HPW}. Clearly, it cannot represent correctly the pion mass dependence of $g_A$ outside a region of extremely small pion masses, as already reported in \cite{HPW}. The ``chiral log'' in the leading-non-analytic term of the  quark mass expansion of $g_A$ is only visible for pion masses well below the physical one \cite{HPW}.
When we analyze the separate contributions to the best fit curve in SSE (the solid/dotted one in Fig.\ref{gauno}) corresponding to the different one-loop diagrams, it turns out that the graph with two delta propagators plays an important role at the pion masses presently manageable on the lattice, in order to counterbalance the trend shown by the graphs with no delta, see Fig.\ref{gadue}. If the $\Delta\,(1232)$ is thought to be too heavy to propagate, the $\Delta \Delta$ graph starts contributing at order $p^5$ in the scheme restricted to pions and nucleons only, via a fifth order tadpole with an attached external axial field. Therefore the contribution from this diagram starts at higher order with respect to our previous calculation in $SU(2)$ BChPT performed up to next-to-leading one-loop.

\begin{figure}[t]
  \begin{center}
    \includegraphics*[width=0.5\textwidth]{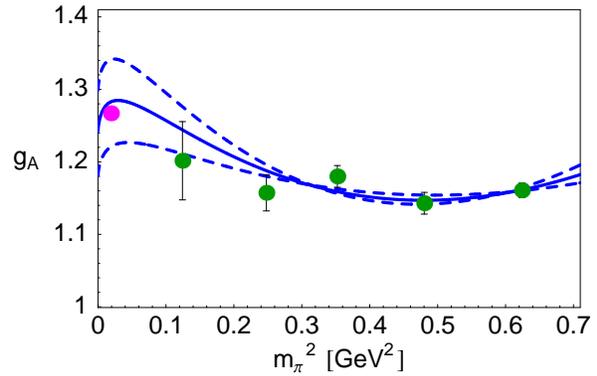}
    \caption{Chiral extrapolation of lattice data by the LHP collaboration \cite{LHP} based on the leading-one-loop expression for $g_A(m_\pi)$ in the scheme with explicit $\Delta\,(1232)$ degrees of freedom.}
    \label{gaextra}
  \end{center}
\end{figure}

Fig.\ref{gaextra} shows a chiral extrapolation based on the leading-one-loop expression in SSE. Here we have analyzed very recent lattice data by the LHP collaboration \cite{LHP}, with two light and one heavier flavors, cf. the talk by Wolfram Schroers in this workshop. The physical point has not been included as input. In the fit an effective low-energy coupling has been constrained according to the analysis by Fettes {\it et al.} of $\pi N \to \pi \pi N$ scattering \cite{fettesuno,fettesdue}, see also \cite{HPW}. The solid and dashed curves correspond to the central, upper and lower value for the previously mentioned coupling, as determined in Ref.\cite{fettesdue}. The axial-$\Delta$-$\Delta$ coupling comes out close to the $SU(4)$ quark model prediction. Interestingly, the inclusion of a heavier quark in the simulations yields results that are statistically compatible with the two-flavor case \cite{PHWga}.

\end{document}